\documentclass[a4paper,
superscriptaddress,
notitlepage,
amsmath,amssymb,
aps,
 onecolumn,
 prx
 floatfix]{revtex4-2}
\usepackage{graphicx}
\usepackage{bm}
\usepackage{amsmath}
\usepackage{amsfonts}
\usepackage{amsbsy}
\usepackage{amssymb}
\usepackage[mathscr]{eucal}
\usepackage{color}
\usepackage{soul}
\usepackage{mathtools}
\usepackage{amsmath,blkarray,booktabs,bigstrut}
\usepackage{tikz}
\usetikzlibrary{calc,arrows,decorations.pathreplacing,backgrounds}

\colorlet{mylinkcolor}{blue!66!black!80}
\usepackage[colorlinks=true,linkcolor=mylinkcolor,citecolor=mylinkcolor,filecolor=cyan,urlcolor=mylinkcolor,breaklinks=true]{hyperref}
\usepackage{xcolor}

\begin{document}

\title{Hallmarks of Deception in Asset-Exchange Models}

\author{Kristian Blom}
\email{kristian.blom@mpinat.mpg.de}
\affiliation{Mathematical bioPhysics group, Max Planck Institute for Multidisciplinary Sciences, G\"{o}ttingen 37077, Germany}

\author{Dmitrii E. Makarov}
\email{makarov@cm.utexas.edu}
\affiliation{Department of Chemistry and Oden
Institute for Computational Engineering and Sciences, The University of Texas at Austin, Austin, Texas 78712, United States}

\author{Alja\v{z} Godec}
\email{agodec@mpinat.mpg.de}
\affiliation{Mathematical bioPhysics group, Max Planck Institute for Multidisciplinary Sciences, G\"{o}ttingen 37077, Germany}

\date{\today}

\date{\today }
\begin{abstract}   
We investigate the transient and steady-state dynamics of the
Bennati-Dragulescu-Yakovenko money game in the presence of
probabilistic cheaters, who can misrepresent
their financial status by
claiming to have no money. We derive the steady-state
wealth distribution per player analytically, and show how the presence of hidden
cheaters can be inferred from the relative variance of wealth per
player. In scenarios with a finite number of cheaters amidst an
infinite pool of honest
players, we identify a critical probability of
cheating at which the total wealth owned by the cheaters experiences a
second-order discontinuity. Below this point, the transition probability to lose money is larger than the probability to gain; conversely, above this point, the direction is reversed. We further establish a threshold
cheating probability at which cheaters collectively possess half of
the total wealth in the game. Lastly, we provide bounds on
the rate at which both cheaters and honest players can gain or lose
wealth, contributing to a deeper understanding of deception in asset
exchange models.  
\end{abstract}
\maketitle
\section{Introduction}\label{SecI}
Asset exchange models, including the Bennati-Dragulescu-Yakovenko (BDY) game \cite{RevModPhys.81.1703, bennati1988metodo, dragulescu2000statistical} and various other asset exchange games \cite{ispolatov1998wealth, PhysRevE.104.014151, greenberg2024twenty, doi:10.1098/rsta.2021.0167, Sinha_2003, angle1986surplus}, have emerged in econophysics as simple, often analytically solvable statistical mechanics models that nevertheless capture certain general features of wealth dynamics in economics. They, in particular, offer an explanation of the fact that wealth distributions are often close to the Boltzmann law \cite{RevModPhys.81.1703, lanchier2017rigorous, lanchier2018rigorous, lanchier2019rigorous, lanchier2024distribution, Cao_2024}, exhibiting characteristics consistent with both positive and negative temperatures \cite{lucente2023randomexchangedynamicsbounds}. More general money exchange models further predict Pareto's power-law wealth distributions and the winner-takes-all economics scenarios \cite{10.1088/1751-8121/ad369b, matthes2008steady, doi:10.1142/S0129183114410083, PhysRevE.89.042804, BOGHOSIAN201715, aydiner2023universal}. These models draw on the analogy between money exchange amongst ``agents'' in economics and energy exchange between particles in physics, often leading to fruitful insights. For example, Boltzmann-like wealth distributions in economics can be linked to the time-reversal symmetry of money exchanges \cite{RevModPhys.81.1703}, which, in turn, is equivalent to fairness of the exchange. Violation of the time-reversal symmetry, which amounts to unfair exchange in economics terms, leads to non-Boltzmann wealth distributions \cite{10.1088/1751-8121/ad369b, FeiCao2023,
  SCAFETTA2004338, Cao_Cortez_2024}. 

In the simplest version of the time-reversible (i.e., fair) game \cite{RevModPhys.81.1703, bennati1988metodo, dragulescu2000statistical}, the wealth $m_i$ of a single player undergoes a one-dimensional random walk. Because the exchange is fair, one may surmise that this walk may be 
unbiased, but this is not the case: the condition that each player's wealth cannot be negative ($m_i\ge 0$) introduces a bias where $m_i$ (assuming $m_i\ne 0$) is less likely to increase than to decrease because of the possibility that the other exchange partner has zero wealth and thus must receive money. This ensures the existence of a steady-state distribution, which can further be shown to be exponential \cite{10.1088/1751-8121/ad369b}. Extensions of this model with unfair game rules were also shown to result in non-Boltzmann wealth statistics such as the Pareto law (a power law) \cite{10.1088/1751-8121/ad369b}.

Here, we extend this model by introducing probabilistic ``cheaters'' into the game. The cheaters misrepresent their financial status with a given probability. Specifically, they can deceive their exchange partners by claiming that they have no money, enabling them to evade any potential losses during the interaction. Such deceptive strategies are prevalent among humans for tax evasion \cite{PICKHARDT2014147, doi:10.17310/ntj.2007.3.16}, which in the United States is estimated to lead to annual losses of approximately $100$ billion US dollars \cite{PhysRevE.101.032305, doi:10.17310/ntj.2009.4.07}. Recently, the need for understanding and quantifying the degree and mechanisms of cheating along with quantifying the ensuing gains has also arisen in an entirely different context: it was discovered that  large language models are also capable of cheating \cite{doi:10.1073/pnas.2317967121, hendrycks2023overviewcatastrophicairisks, azaria2023internal}. Deceptive strategies can also be observed outside the realm of humans, among various species in the animal and plant kingdoms \cite{Sekrst2022-EKRELD, rowell2006animals}.
General aspects of deception are well-studied in the context of (evolutionary) game
theory \cite{KAJACKAITE2017433, vanDitmarsch2012,
  10.1007/978-3-030-32430-8_5,
  rubinstein2015honest, szamado2000cheating, rowell2006animals, math12030414}, where particular focus is made on the incentive behind truth-telling and lying \cite{7ac0d2e3-1bb7-3732-b97c-b5f65bd5f784, PhysRevE.101.032305, doi:10.1098/rsif.2019.0211}. However, the resulting ramifications of lying in a many-player context remain largely unexplored.  Consequently, it is crucial to investigate both the mechanisms of deception and detection methods within such many-agent environment. 
  In a system that consists of both honest players and cheaters, we investigate how cheating alters
  the transient dynamics as well as steady-state distributions
of wealth of
the respective types of players.  

\section{Outline and summary of main results}
The paper is structured as follows:~In Section~\ref{SecII} we provide
a detailed description of the model and introduce the concept of
probabilistic cheaters within the framework of the
BDY game.

Section~\ref{SecIII} focuses on the
steady-state probabilities, denoted as $\pi^{c}(m)$ and $\pi^{h}(m)$,
representing the probability that a single cheater or honest player
possesses an amount of money $m$, respectively. Based on a recently developed mean-field approximation \cite{10.1088/1751-8121/ad369b} we derive the wealth distribution for the cheaters and honest players analytically, and the results are given by Eqs.~\eqref{solc}-\eqref{pifm}. Our analysis reveals a critical
scaling relationship between the probability of cheating $q_{c}$ and
the average wealth per person, $\langle m \rangle$, setting a
threshold at which cheaters begin to benefit from being
deceptive. We further demonstrate in Section~\ref{SecIII.F} how the presence of hidden
cheaters can be deduced from the relative variance of the overall
steady-state probabilities. Thereby, an external observer can detect the presence of hidden cheaters without requiring any additional knowledge besides the number of players and total amount of money. 

Next, in Section~\ref{SecIV}, we examine
the total wealth possessed by cheaters and honest players, denoted by $\langle M_{c} \rangle$ and $\langle M_{h} \rangle$, respectively. We
identify a threshold cheating probability, $q_{1/2}$ (see Eq.~\eqref{qhalf}), at which the pool of cheaters and honest players have an equal amount of money, i.e., $\langle M_{c} \rangle = \langle M_{h} \rangle$. Additionally,
we establish the existence of a critical cheating probability, $q_{\rm
  crit}$ (see Eq.~\eqref{qcrit}), at which the wealth of a finite pool of cheaters interacting
with an infinite pool of honest players undergoes a second-order
discontinuity. For $q_{c}<q_{\rm crit}$ we find that $\langle M_{c} \rangle=0$, whereas for $q_{c}>q_{\rm crit}$ we have $\langle M_{c} \rangle >0$. Finally, in Section~\ref{SecV.D} we analyze the transient dynamics of wealth
accumulation for both cheaters and honest players, presenting bounds
on how quickly each party can gain or lose money
over time. 

We conclude in Section~\ref{SecV} by summarizing the main
results, and provide an outlook for future work. Details of calculations and proofs are given in Appendices \ref{AppendixA}-\ref{AppendixD}. 

\section{Model}\label{SecII}
\subsection{Money exchange rules}\label{SecIIA}
\begin{figure}
    \centering
    \includegraphics[width=\textwidth]{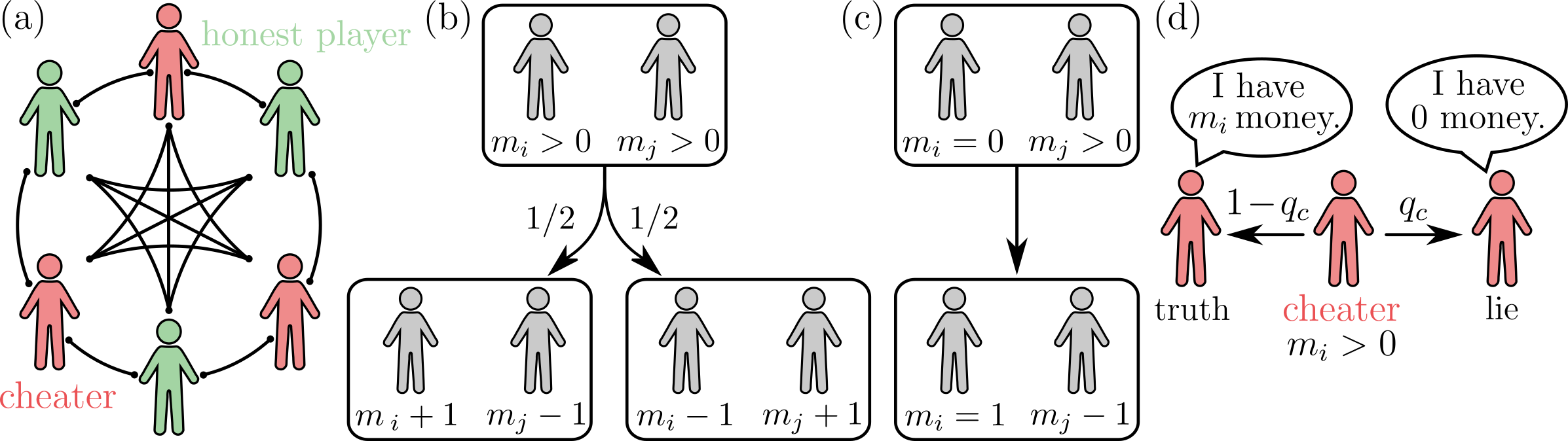}
    \caption{(a) The money game, including interactions between honest
      players (green) and cheaters (red), who engage in stochastic
      money exchanges.~(b) During interactions between two players
      with nonzero money, the exchange occurs with equal probability
      in either direction.~(c)~If one player has zero money while the
      other has a nonzero amount, the exchange will only occur from
      the player with nonzero money to the player with zero money.~(d)
      A cheater is characterized by a probability $q_{c}$ of being deceptive
      when encountering another player. Once deciding to
      deceive, a cheater falsely claims to have $0$ money, thus gaining money
      if the opposing player possesses a nonzero amount.  
    } 
    \label{Fig1}
\end{figure}
Following refs.~\cite{RevModPhys.81.1703,
  dragulescu2000statistical} we consider a
pool of $N$ players, where each player $i\in\{1,2,...,N\}$ has
$m_i$ units of money (see Fig.~\ref{Fig1}a). The total amount of money,
\begin{equation}
M = \sum_{i=1}^N m_i,
\end{equation}
remains constant and is determined by the initial conditions of the
game. In each interaction, a pair of players, denoted as $i$ and $j$,
is selected at random. If each player has a positive amount of
money, i.e., $m_i > 0$ and $m_j > 0$, they engage in an exchange of
money,
described mathematically by: 
\begin{equation}
    m_i \rightarrow m_i \pm 1, \quad m_j \rightarrow m_j \mp 1,
\end{equation}
where the ``$+$'' and ``$-$'' signs indicate the direction of the
exchange. In this work, we will only be interested in the unbiased
money game, where the signs ``$+$" and ``$-$" are selected with the same
probability (see Fig.~\ref{Fig1}b; for studies of biased
money games see \cite{10.1088/1751-8121/ad369b, FeiCao2023,
  SCAFETTA2004338,Cao_Cortez_2024}). When one of the players has no money, only one direction of exchange is possible, and 
the exchange becomes deterministic. For example, if $m_i=0$ and $m_j >
0$, the result of the exchange is $m_i \rightarrow 1$, $m_j
\rightarrow m_j-1$ (see Fig.~\ref{Fig1}c). If $m_i=0$ and $m_j=0$,
then no exchange takes place, i.e., $m_i \rightarrow m_i$, $m_j
\rightarrow m_j$.  

In the absence of cheaters, the properties of this model are well known. In particular, for $N\gg 1$, this model results in Boltzmann wealth statistics. The Boltzmann distribution is often a good approximation for real-world distributions of wealth \cite{RevModPhys.81.1703}. Other scenarios that have analogs in economics can also arise in this model when unfair exchange rules are implemented. In particular, the model predicts a power-law distribution of wealth (Pareto law) in certain regimes, which provides a more accurate representation of the tails in real-world wealth distributions \cite{RevModPhys.81.1703}.  In this work, we focus our attention only on the simplest case with fair exchanges (in the absence of cheaters).
\subsection{Introducing probabilistic cheaters}\label{SecIIB} 
Building upon the unbiased money game, we now incorporate a subset of
probabilistic cheaters, which do not exactly follow the above exchange rules, as illustrated in Fig.~\ref{Fig1}d. Whenever a
cheater encounters another player, said cheater will deceive the other
player with probability $q_{c}
\in [0,1]$ by claiming to have
no money. Consequently, if the cheater opts
to lie and the opposing player has nonzero wealth, the cheater
automatically receives money. This tactic may be classified as a
``black lie", as it solely benefits the liar
\cite{7ac0d2e3-1bb7-3732-b97c-b5f65bd5f784}. 
In instances where the
opposing player has no money, the cheating attempt is ineffective,
resulting in no exchange of money. 
\begin{figure}
    \centering
    \includegraphics[width=\textwidth]{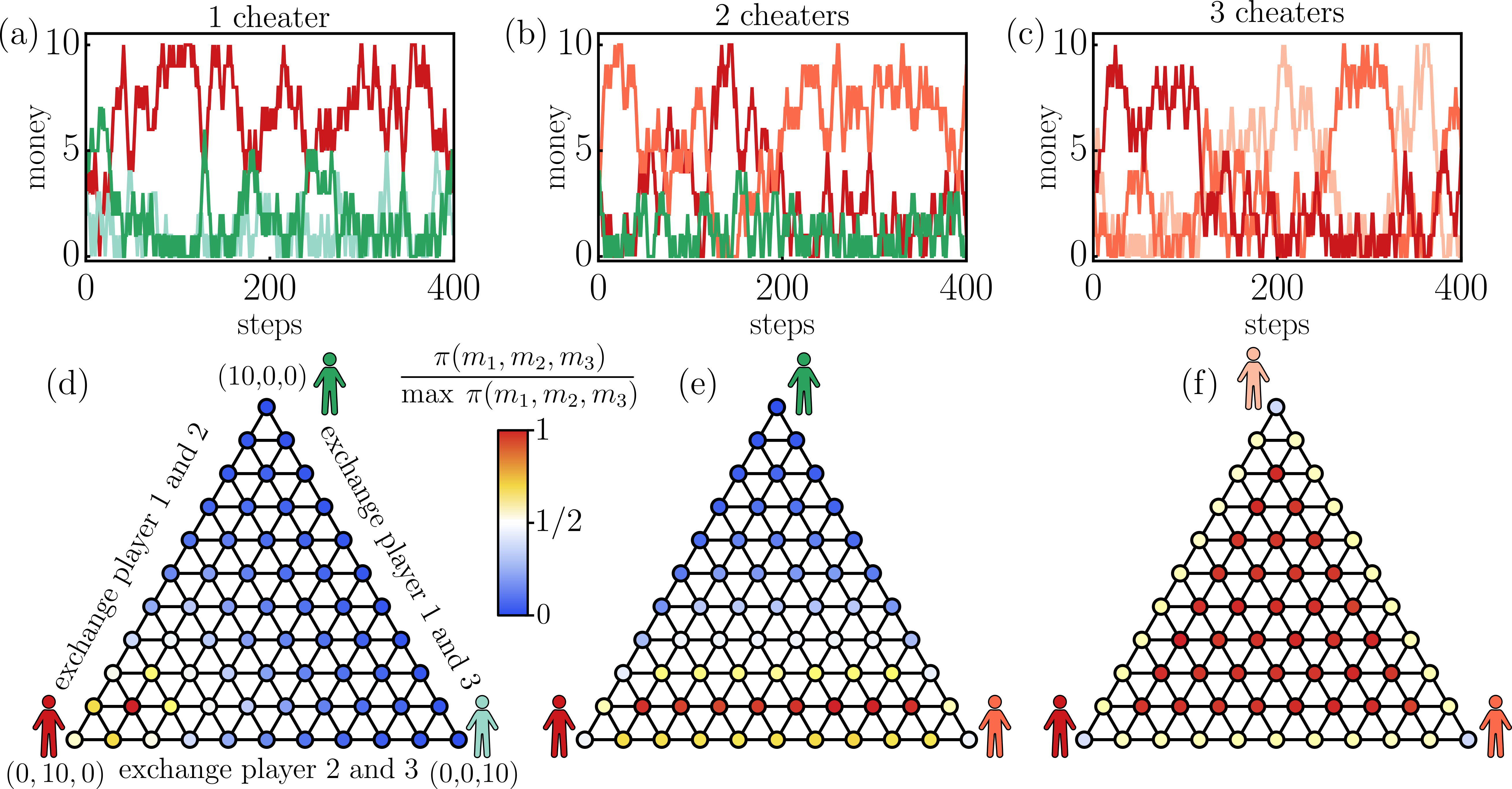}
    \caption{Stochastic trajectories (a-c) and steady-state probabilities (d-f) for the money game involving $N{=}3$ players and $M{=}10$ units of money, obtained with Monte-Carlo simulations. From left to right, we increment the number of cheaters (red players) in the game from $1$ (a,d) to $3$ (c,f). The cheating probability is maintained at $q_{c}=0.2$. In panels (d-f), each colored node indicates the steady-state probability for a specific configuration of the system, which can be written as a triplet $(m_{1},m_{2},m_{3})$.}
    \label{Fig2}
\end{figure}
\subsection{Evolution equation and steady-state probability}
Let $N_{c}$ be the number of cheaters and $N_{h}$ the number of honest players, such that $N=N_{c}+N_{h}$. Furthermore, let $n\in \mathbb{N}$ denote the total number of encounters between the players. The probability $P(\mathbf{m}_{c},\mathbf{m}_{h};n)\in[0,1]$ to have a specific configuration of money among the cheaters $\mathbf{m}_{c}=\{m^{c}_{1},...,m^{c}_{N_{c}}\}$ and honest players $\mathbf{m}_{h}=\{m^{h}_{1},...,m^{h}_{N_{h}}\}$ after $n$ encounters obeys the evolution equation
\begin{equation}
     \mathbf{P}(\mathbf{m}_{c},\mathbf{m}_{h};n+1)=\mathbf{Q}\mathbf{P}(\mathbf{m}_{c},\mathbf{m}_{h};n), 
\end{equation}
where $\mathbf{P}(\mathbf{m}_{c},\mathbf{m}_{h};n)$ is a vector of size $|\Omega|\equiv \binom{M+N-1}{N-1}$ that
contains the probabilities $P(\mathbf{m}_{c},\mathbf{m}_{h};n)$ for
each possible configuration after $n$ encounters. The transition
matrix $\mathbf{Q}$ of size $|\Omega|\times|\Omega|$ contains elements
$Q_{ij}$ that represent the transition probabilities from state $j$ to
$i$. These transitions are determined by the rules defined in
Secs.~\ref{SecIIA} and \ref{SecIIB}. Note that the diagonal elements
of $\mathbf{Q}$ are given by $Q_{ii}=1-\sum_{j\neq i}Q_{ji}$, such
that the probability is conserved. 

Since the Markov chain associated with $\mathbf{Q}$ is
aperiodic and irreducible for $0 \leq q_c < 1$
\cite{garibaldi2013interplay}, there exists a unique steady-state
probability $\boldsymbol{\pi}(\mathbf{m}_{c}, \mathbf{m}_{h}) \equiv
\lim_{n \rightarrow \infty} \mathbf{P}(\mathbf{m}_{c}, \mathbf{m}_{h};
n)$. This steady-state probability corresponds to the eigenvector of
$\mathbf{Q}$ with an eigenvalue of $\lambda = 1$ \cite{norris1998markov},
i.e., 
\begin{equation}
    (\mathbf{Q}-\mathbf{1})\boldsymbol{\pi}(\mathbf{m}_{c},\mathbf{m}_{h})=\mathbf{0},
    \label{steadystate}
\end{equation}
where $\mathbf{1}$ is the identity matrix, and $\mathbf{0}$ is the zero vector. In Fig.~\ref{Fig2} the steady-state probability for a three-player game with $M=10$ and varying number of cheaters is shown, based
on Monte-Carlo simulations. In the case of a single cheater, we can
clearly see that a relatively small cheating probability of $q_c=0.2$
can already lead to substantial gains for the cheater (see
Fig.~\ref{Fig2}d), as the largest steady-state probability peaks
around $8$ money units for the cheater and $1$ money unit for each of
the honest players.   

Interestingly, despite the presence of cheaters that bias the transition rates, the transition matrix still satisfies the detailed balance condition. That is, for every pair of states $i$ and $j$ we have:
\begin{equation} 
\pi_{j}Q_{ij}=\pi_{i}Q_{ji}. 
\label{DB}
\end{equation}
As a result, the Markov chain is time-reversible. In Appendix \ref{AppendixA}, we apply Kolmogorov's cycle criterion to
demonstrate that this holds in general, regardless of the number
of cheaters and honest players involved
\cite{kelly2011reversibility,kolmogoroff1936theorie}. 
\section{Steady-state probability distribution of wealth for a single player}\label{SecIII}
The factorial growth of the state space with increasing $N$ and/or $M$
renders 
Eq.~\eqref{steadystate}
analytically intractable, such that it is not possible to compute the steady-state
probability. To circumvent this combinatorial complexity, we focus
instead on the steady-state probability of wealth for a single cheater
or honest player, denoted by $\pi^{c}(m)$ and $\pi^{h}(m)$
respectively. Formally, these steady-state probabilities are
projections of the full steady-state probability: 
\begin{align}
    \pi^{c}(m)&\equiv\frac{1}{N_{c}} \sum_{\mathbf{m}_{c}}\sum_{\mathbf{m}_{h}}\sum_{i=1}^{N_{c}}\pi(\mathbf{m}_{c},\mathbf{m}_{h})\delta_{m^{c}_{i},m}, \nonumber \\
    \pi^{h}(m)&\equiv\frac{1}{N_h} \sum_{\mathbf{m}_{c}}\sum_{\mathbf{m}_{h}}\sum_{i=1}^{N_{h}}\pi(\mathbf{m}_{c},\mathbf{m}_{h})\delta_{m^{h}_{i},m},
\end{align}
where $\delta_{x,y}=1$ if $x=y$ and $0$ elsewhere. In Appendix \ref{AppendixB} we derive $\pi^{c}(m)$ and $\pi^{h}(m)$ in the
macroscopic limit, which is defined as 
\begin{align}
    \lim\nolimits^{M\rightarrow\infty}_{N\rightarrow\infty}[M/N]&\equiv \langle m \rangle \in \mathbb{R}_{>0} , \nonumber \\  \lim\nolimits^{N_{h}\rightarrow\infty}_{N\rightarrow\infty}[N_{h}/N]&\equiv \Phi_{h} \in [0,1], \\
    \lim\nolimits^{N_{c}\rightarrow\infty}_{N\rightarrow\infty}[N_{c}/N]&\equiv \Phi_{c} \in [0,1], \nonumber
\end{align}
where $\Phi_{h}$ and $\Phi_{c}$ are the number densities of honest players
and cheaters, respectively, which obey
$\Phi_{h}+\Phi_{c}=1$. Employing the mean-field approximation
\cite{10.1088/1751-8121/ad369b} (which becomes exact in the
macroscopic limit since all players interact with each other,
i.e.\ we have an ``all-to-all'' interaction)
we obtain analytical expressions for  $\pi^{c}(m)$ and $\pi^{h}(m)$, 
which are given in the next sections.
\subsection{Steady-state probability distribution of wealth for a cheater}
\begin{figure}
    \centering
    \includegraphics[width=\textwidth]{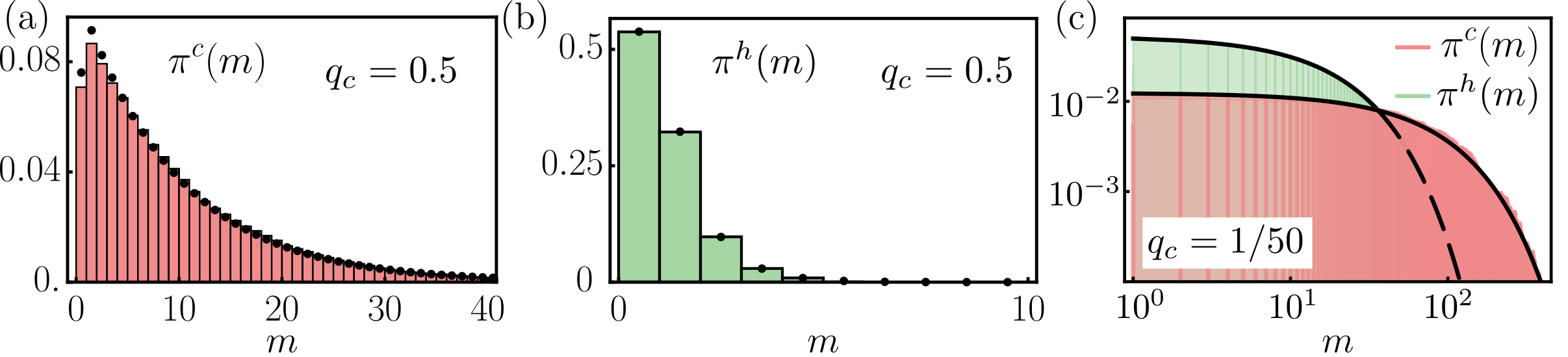}
    \caption{(a-c) Steady-state probabilities of wealth for a single cheater (a,c) and honest player (b,c) in the money game. Colored bars are obtained with Monte-Carlo simulations with $N_{c}=10$ cheaters and $N_{h}=10$ honest players. In (a,b) we set $M=100$ money units and cheating probability $q_c=0.5$. In (c) we set $M=1000$ and $q_{c}=1/50$. Black dots in (a,b) correspond to the analytical results given by Eqs.~\eqref{pic2} and \eqref{pifm}, respectively. Black lines in (c) correspond to the asymptotic exponential distribution given by Eq.~\eqref{intermediate2}.}
    \label{Fig3}
\end{figure}

The  steady-state probability distribution of wealth for a single cheater in the macroscopic limit reads
\begin{align}
    \pi^{c}(0)&=\frac{\pi^{h}(0)-q_{c}}{1-q_{c}}, \label{solc} \\
    \pi^{c}(m) &= \frac{2(\pi^{h}(0)-q_c)}{1-q^2_{c}}\left(\frac{1-\pi^{h}(0)}{1+\pi^{h}(0)}\frac{1+q_c}{1-q_c}\right)^{m}, \ m\geq 1, \label{pic2}
\end{align}
where $\pi^{c}(0)$ is the probability for a cheater to have zero
money, and similarly $\pi^{h}(0)$ the probability for an honest player to
have zero money (defined in Eq.~\eqref{solf}). Notably,
Eq.~\eqref{pic2} is remarkably accurate even for finite systems, as
demonstrated in Fig.~\ref{Fig3}a, where we compare the results from
Eq.~\eqref{pic2} (indicated by the black dots) with finite-size
Monte-Carlo simulations (represented by the colored bars). 

Looking at Eq.~\eqref{solc}, we note that the inequality $0 \leq \pi^{c}(0) \leq 1$
leads to the conclusion that $q_{c} \leq
\pi^{h}(0) \leq 1$. This implies that the steady-state probability of
an honest player having zero money is 
bounded from below by the
cheating probability in the presence of cheaters. This finding will
have significant implications in the case of a small pool
of cheaters in Sec.~\ref{SecIV.B}.
\subsection{Steady-state probability  of wealth for an honest player}
The steady-state probability for an honest player in the range $0<\Phi_h\leq1$ reads,
\begin{align}
    \pi^{h}(0)&=2\sqrt{\frac{\Delta_0}{3}}\cos{\left(\frac{1}{3}\arccos{\left(-\frac{3\Delta_1}{2\Delta_0}\sqrt{\frac{3}{\Delta_0}}\right)}\right)}+\frac{1}{3}(\Phi_{h}q_c-2\langle m \rangle), \label{solf} \\
    \pi^{h}(m)&=2\pi^{h}(0)\left(\frac{1-\pi^{h}(0)}{1+\pi^{h}(0)}\right)^{m}, \ m\geq 1,
     \label{pifm}
\end{align}
which is shown in Fig.~\ref{Fig3}b together with results obtained from
Monte-Carlo simulations. The specific form of $\pi^{h}(0)$ comes from
the solution of a cubic equation (see Eq.~\eqref{cond2}), where
the auxiliary functions $\Delta_0\geq 1$ and $\Delta_1$ are  
\begin{align}
     \Delta_{0}&\equiv 1+2\langle m \rangle q_{c}+(1/3)(2\langle m \rangle -\Phi_{h}q_{c})^2, \nonumber \\ 
      \Delta_{1}&\equiv (2/27)(2\langle m \rangle -\Phi_{h}q_{c})^3+(1/3)(2\langle m \rangle -\Phi_{h}q_{c})(1+2\langle m \rangle q_{c})+\Phi_hq_c. \label{del1}
\end{align}
Note that for $\Phi_{h}=0$ Eq.~\eqref{pifm} does not have any meaning as there are no honest players in the game, yet we still can insert the final expression into Eqs.~\eqref{solc}-\eqref{pic2} to obtain the steady-state of the cheaters in this scenario. 

To grasp how Eqs.~\eqref{solc}-\eqref{pifm} vary for different parameter settings, we analyze the steady-state probabilities for different scenarios in the next three sections.  
\subsection{Scenario 1: All honest; $q_{c}=0$ or $\Phi_{h}=1$.}\label{SecIII.C}
For $q_{c}=0$ or $\Phi_{h}=1$, the game consists solely of honest players. Setting $q_c=0$ in Eq.~\eqref{solf}, we find that $\pi^{h}(0)$ simplifies to the original result reported in \cite{10.1088/1751-8121/ad369b}:
 \begin{equation}
     \pi^{h}(0)=\sqrt{\langle m \rangle ^{2}+1}-\langle m \rangle .
     \label{sol2}
 \end{equation}
 This probability will become relevant in Sec.~\ref{SecIV.B}, where we identify $\pi^{h}(0)$ with a critical cheating probability. Additional analysis of this scenario can be found in \cite{10.1088/1751-8121/ad369b}. Furthermore, in Appendix \ref{AppendixC} we prove that Eq.~\eqref{sol2} is a lower bound on $\pi^{h}(0)$ whenever there are cheaters present in the game. This is expected, as the presence of cheaters makes it more likely to lose money. 
\subsection{Scenario 2: All cheating; $0<q_{c}<1$ and $\Phi_{h}=0$.}
When $0<q_{c}<1$ and $\Phi_{h}=0$, the game solely consists of the cheaters, and Eq.~\eqref{solc} simplifies to
 \begin{equation}
     \pi^{c}(0)=\frac{\sqrt{\langle m \rangle ^{2}+2\langle m \rangle q_{c}+1}-\langle m \rangle - 1}{1-q_{c}}+1.
     \label{sol3}
 \end{equation}
For $q_{c}=0$, Eq.~\eqref{sol3} further simplifies to
Eq.~\eqref{sol2}. Furthermore, we observe that $\pi^{c}(0)$ is a
strictly positive increasing function with increasing
$q_{c}$. Consequently, $\pi^{c}(0)>\sqrt{\langle m \rangle ^{2}{+}1}{-}\langle m \rangle$  for $0<q_{c}\leq1$,
and therefore being an honest player among honest
players is preferable to being a cheater among fellow cheaters, as the former
reduces the probability of having no money. Note that for $q_{c}=1$
the dynamics is \textit{frozen}, as every cheater declares to have
zero money, and therefore the system becomes non-ergodic.  
\subsection{Scenario 3: Mixture of cheaters and honest players in the $\langle m \rangle \rightarrow \infty$ limit}\label{SecIII.E}
Finally, we are interested in the limit where the average wealth per player goes to infinity, i.e., $\langle m \rangle \rightarrow \infty$. When taking this limit, we need to consider the scaling of the cheating probability $q_c$ with $\langle m \rangle$, as the result is different depending on whether $q_c \langle m \rangle \rightarrow 0$, $q_c \langle m \rangle \rightarrow {\rm constant}$, or $q_c \langle m \rangle \rightarrow \infty$, leading to  three different regimes:
\\
\\
\textit{1. The weak-cheating regime} $\langle m \rangle \rightarrow \infty$ \textit{and} $q_c \langle m \rangle \rightarrow 0$: 
In this limit, we take $\langle m \rangle \rightarrow \infty$ while keeping $q_{c}=\mathcal{A}_{c}/\langle m \rangle^{\mathcal{B}}$ with $\mathcal{B}>1$ and $\mathcal{A}_{c}={\rm constant} \in \mathbb{R}_{>0}$. 
Then, Eq.~\eqref{solf} attains the following asymptotic solution
\begin{align}
    \pi^{h}(0)=\pi^{c}(0)= \frac{1}{2\langle m \rangle}+\mathcal{O}\left(\frac{1}{\langle m \rangle^{\mathcal{B}}}\right),
\end{align}
which agrees with the findings for an honest-player game (see \cite{10.1088/1751-8121/ad369b} for a comprehensive analysis). Notably, there is no distinction between the steady-state probabilities of cheaters and honest players.
\\
\\
\textit{2. The intermediate-cheating regime} $\langle m \rangle \rightarrow \infty$ \textit{and} $q_c \langle m \rangle = {\rm constant}$:
In this limit we simultaneously take  $\langle m \rangle \rightarrow \infty$ and $q_c \rightarrow 0$ while keeping $q_{c}=\mathcal{A}_{c}/\langle m \rangle$ with $\mathcal{A}_{c}={\rm constant} \in \mathbb{R}_{>0}$.  Then, Eqs.~\eqref{solc} and \eqref{solf} attain the asymptotic solution
\begin{align}
    \pi^{c}(0)&= \frac{1-2\mathcal{A}_{c}+\sqrt{1+4\mathcal{A}_{c}(1+\mathcal{A}_{c}-2\Phi_{h})}}{4\langle m \rangle}+\mathcal{O}\left(\frac{1}{\langle m \rangle^{3}}\right) \nonumber , \\
    \pi^{h}(0)&= \frac{1+2\mathcal{A}_{c}+\sqrt{1+4\mathcal{A}_{c}(1+\mathcal{A}_{c}-2\Phi_{h})}}{4\langle m \rangle}+\mathcal{O}\left(\frac{1}{\langle m \rangle^{3}}\right) \label{p0int2},
\end{align}
and upon inserting  back into Eqs.~\eqref{pic2} and \eqref{pifm} we obtain the exponential distribution (note that $\simeq$ means asymptotic equality, i.e., $A\simeq B$ when $\lim_{\langle m \rangle \rightarrow \infty} A/B=1$)
\begin{align}
    \pi^{c}(m)&\simeq 2\pi^{c}(0)\exp{\left(-2\pi^{c}(0)m\right)}, \ m\geq1, \nonumber \\
    \pi^{h}(m)&\simeq2\pi^{h}(0)\exp{(-2\pi^{h}(0)m)}, \ m \geq 1. \label{intermediate2}
\end{align}
Based on the above expressions, we observe that even small
cheating probabilities, on the order of $q_{c} \propto 1/\langle m
\rangle$, can lead to significant advantages for the cheaters. This is
illustrated in Fig.~\ref{Fig3}c for a game with $\langle m \rangle =
50$, $\Phi_{h} = 1/2$, and $q_{c} = 1/50$ (hence $\mathcal{A}_{c} =
1$). Remarkably, the results show a very good agreement with Monte-Carlo
simulations, indicating that the asymptotic solution converges quickly
for large $\langle m \rangle$. This convergence is further evidenced
by the remainder term in Eqs.~\eqref{p0int2}, which is
on the order of $\langle m \rangle^{-3}$. 
\\
\\
\textit{3. The strong-cheating regime}, $\langle m \rangle \rightarrow \infty$ \textit{and} $q_c \langle m \rangle \rightarrow \infty$:
This scaling limit is obtained by setting
$q_{c}=\mathcal{A}_{c}/\langle m \rangle^{\mathcal{B}}$ with
$\mathcal{B}<1$ and $\mathcal{A}_{c}={\rm constant} \in
\mathbb{R}_{>0}$. In particular, this includes setting $q_{c}={\rm
  constant}>0$. In this regime the asymptotic solutions of Eq.~\eqref{solc} and Eq.~\eqref{solf} read
 \begin{align}
     \pi^{c}(0)&= \frac{\Phi_{c}(1+q_{c})}{2\langle m \rangle}+\mathcal{O}\left(\frac{1}{\langle m \rangle^{2}}\right), \nonumber \\ 
    \pi^{h}(0)&= q_c+\frac{\Phi_{c}(1-q^{2}_{c})}{2\langle m \rangle}+\mathcal{O}\left(\frac{1}{\langle m \rangle^{2}}\right). 
 \end{align}
 Upon inserting into Eqs.~\eqref{pic2} and \eqref{pifm}, we obtain the steady-state probabilities
 \begin{align}
    \pi^{c}(m)&\simeq\frac{\Phi_{c}}{\langle m \rangle }\exp{\left(-\frac{\Phi_{c}}{\langle m \rangle}m\right)} , \ m\geq1, \nonumber \\
    \pi^{h}(m)&\simeq
    2q_c \left(\frac{1-q_{c}}{1+q_{c}}\right)^{m}, \ m \geq 1.
\end{align}
In this regime, the cheaters completely dominate the game, and
consequently own the majority of money. Interestingly, the steady
state for the cheaters is independent of the cheating probability $q_{c}$,
whereas for the honest players it is independent of the number
density of honest players $\Phi_{h}$.
\subsection{Detection of hidden cheaters from the steady-state distribution}\label{SecIII.F}
Consider now an external observer who is unable to differentiate
between cheaters and honest players. Upon sampling the steady-state
distribution of wealth per player, this external observer will observe
the distribution
\begin{equation}
    \pi(m)=\Phi_{h}\pi^{h}(m)+\Phi_{c}\pi^{c}(m).
\end{equation}
To see how the presence of hidden cheaters affects the total
steady-state distribution, we inspect the variance, ${\rm var}(m)\equiv \langle m^{2} \rangle {-} \langle m \rangle^{2}$, which reads
\begin{equation}
    {\rm var}(m)= \langle m \rangle \left(\frac{1}{\pi^{h}(0)}+\frac{(1{-}[\pi^{h}(0)]^{2})\Phi_cq_c}{(\pi^{h}(0){-}q_{c})(\pi^{h}(0)-\Phi_h q_c)} - \langle m \rangle \right),
    \label{varm}
\end{equation}
with $\pi^{h}(0)$ given by Eq.~\eqref{solf}. 
In Appendix \ref{AppendixC} we prove that when hidden cheaters are present with parameters $0<q_{c}\leq 1$ and $0<\Phi_{c}\leq1$, the following inequality for the relative variance holds
\begin{equation}
    \frac{{\rm var}(m)}{\langle m \rangle^{2}}> \frac{{\rm var}(m)|_{q_{c}=0}}{\langle m \rangle^{2}} = \sqrt{1+1/\langle m \rangle^{2}}, \ {\rm for} \ \langle m \rangle >0.
    \label{varineq}
\end{equation}
Hence, an external observer can detect hidden cheaters by determining whether the relative variance of the steady-state distribution obeys Eq.~\eqref{varineq}.
In the large-money-per-player limit, Eq.~\eqref{varineq} takes the simple form
\begin{equation}
    \lim_{\langle m \rangle \rightarrow \infty} \frac{{\rm var}(m)}{\langle m \rangle^{2}}>1+\mathcal{O}\left(\frac{1}{\langle m \rangle^{2}}\right),
    \label{simplebound}
\end{equation}
which is a consequence of the addition
of two exponential distributions with different parameters, namely that of the cheaters and honest players. For example, if we consider
the intermediate-cheating regime where $\pi^{h}(0)$ is given by
Eq.~\eqref{p0int2}, the relative variance
reads (recall that $q_{c}=\mathcal{A}_{c}/\langle m \rangle$ with $\mathcal{A}_{c}>0$ and $0<\Phi_{h}\leq1$)
\begin{align}
    \lim_{\langle m \rangle \rightarrow \infty}\frac{{\rm var}(m)}{\langle m \rangle^{2}}&= -1+\frac{4+16\mathcal{A}_{c}(1+\mathcal{A}_{c}-2\Phi_h)}{1+4\mathcal{A}_{c}(1+\mathcal{A}_{c}-2\Phi_{h})+(1+2\mathcal{A}_{c}(1-2\Phi_{h}))\sqrt{1+4\mathcal{A}_{c}(1+\mathcal{A}_{c}-2\Phi_{h})}}+\mathcal{O}\left(\frac{1}{\langle m \rangle^{2}}\right)  \nonumber \\
    &> -1+\frac{4+16\mathcal{A}_{c}(1+\mathcal{A}_{c})}{1+4\mathcal{A}_{c}(1+\mathcal{A}_{c})+(1+2\mathcal{A}_{c})\sqrt{1+4\mathcal{A}_{c}(1+\mathcal{A}_{c})}}+\mathcal{O}\left(\frac{1}{\langle m \rangle^{2}}\right) \nonumber \\
    &=1+\mathcal{O}\left(\frac{1}{\langle m \rangle^{2}}\right),
\end{align}
where in the first line we used that the expression on the right-hand side is a monotonically increasing function of $\Phi_{h}$ for $0\leq\Phi_h\leq1$, and therefore attains a minimum value at $\Phi_{h}=0$ which we inserted in the second line. 

Importantly, to effectively utilize Eqs.~\eqref{varineq}-\eqref{simplebound} for detecting hidden cheaters, it is essential to know beforehand that only honest players and cheaters are participating in the game. For instance, if generous players, who give money according to a specific probability (essentially the opposite of cheating), are involved, the relative variance could still surpass the limits established by Eqs.~\eqref{varineq}-\eqref{simplebound}, even in a scenario without any cheaters.
\section{Average wealth owned by cheaters and honest players}\label{SecIV}
From the perspective of a group of cheaters, the most interesting
quantity  is their total wealth, which can be computed from
Eqs.~\eqref{pic2} and \eqref{pifm}.
Let $\langle M_{c} \rangle$ and $\langle M_{h} \rangle$ be
the total wealth of the cheaters and honest players, respectively, in the
steady-state. Due to conservation of money we have $(\langle M_{c}\rangle+\langle M_{h} \rangle)/M=1$. Then, on average, a 
single cheater or honest player has accumulated a wealth 
 \begin{align}
     \langle m_{c} \rangle \equiv \frac{\langle M_{c} \rangle}{N_{c}}  &=\frac{1-[\pi^{h}(0)]^2}{2(\pi^{h}(0)-q_{c})}, \label{mc} \\
    \langle m_{h} \rangle \equiv \frac{\langle M_{h} \rangle}{N_{h}} &= \frac{1-[\pi^{h}(0)]^2}{2\pi^{h}(0)}, \label{mf}
\end{align}
where $\pi^{h}(0)$ is given by Eq.~\eqref{solf}. Note that Eqs.~\eqref{mc}-\eqref{mf} solve the equation
$\Phi_{c}\langle m_{c} \rangle + \Phi_h\langle m_{h} \rangle=\langle m
\rangle$ upon inserting $\pi^{h}(0)$ (see also Appendix \ref{AppendixB}). The
total fraction of money owned by the cheaters and honest players is
$\Phi_{c}\langle m_{c} \rangle/\langle m \rangle = \langle
M_{c} \rangle/M \in [0,1]$ and $\Phi_{h}\langle m_{h} \rangle/\langle
m \rangle = \langle M_{h} \rangle/M \in [0,1]$, which is shown in
Fig.~\ref{Fig4}a as a function of cheating probability.   
\subsection{When do the cheaters own most of the money?}
\begin{figure}
    \centering
    \includegraphics[width=\textwidth]{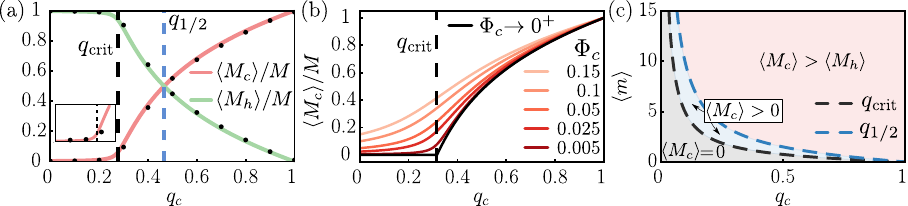}
    \caption{(a) Total fraction of money owned by 
      cheaters (red) and honest players (green) as a function of the
      cheating probability $q_c$.~Dots are obtained with
      Monte-Carlo simulations, while full lines are obtained from
      Eqs.~\eqref{mc} and \eqref{mf}, respectively. We consider the
      following system parameters $(N_{c},N_{h},M)=(1,299,500)$. The blue dashed line correspond to $q_{1/2}$ given by Eq.~\eqref{qhalf}, at which
      the cheaters and honest players have equal wealth. Inset in (a): a
      magnification of the region $[0,0.4]\times[0,0.3]$. (b) Scaling of
      the wealth of cheaters $\langle M_{c} \rangle/M$ with a
      decreasing fraction of cheaters $\Phi_{c}$ for $\langle m
      \rangle = 10/7$. In the limit
      $\Phi_{c}\rightarrow 0^{+}$ (black solid line given by Eq.~\eqref{gain2}) there is a second-order
      discontinuity at $q_{\rm crit}$ (see Eq.~\eqref{qcrit}),
      indicated with the black dashed line. (c) Phase diagram for the total amount of money owned by the cheaters $\langle M_{c} \rangle$ in the low density limit ($\Phi_{c}\rightarrow 0^{+}$)  as a function of the cheating probability $q_{c}$ and the average amount of money per player $\langle m \rangle$. In the gray region $\langle M_{c} \rangle=0$, in the blue region $\langle M_{c} \rangle>0$, and in the red region $\langle M_{c} \rangle >\langle M_{h}\rangle$. The blue dashed line is given by Eq.~\eqref{qhalf} for $\Phi_h=1$, and the black dashed line is given by Eq.~\eqref{qcrit}.} 
    \label{Fig4}
\end{figure}

By analyzing Eqs.~\eqref{mc} and \eqref{mf}, we find that the total
wealth of the cheaters in the steady-state surpasses that of the
honest players, i.e., $\langle M_{c} \rangle > \langle M_{h} \rangle$,
whenever the following condition is met
\begin{equation}
    q_{c}>\pi^{h}(0)(1-\Phi_{c}/\Phi_{h}),
\end{equation}
where we recall that $\pi^{h}(0)$ is given by Eq.~\eqref{solf} and explicitly depends on $q_{c}$.
Note that for $\Phi_{c}\geq\Phi_{h}$ the right-hand side becomes negative, and therefore the cheaters will collectively always have more (or equal) money than the honest players. However, for $0<\Phi_{c}<\Phi_{h}$ there is a threshold cheating probability above which $\langle M_{c} \rangle>\langle M_{h} \rangle$, which reads
\begin{equation}
    q_{1/2}\equiv\frac{(\Phi_{h}-1/2)([\langle m \rangle^{2}+4\Phi^{2}_{h}]^{1/2}-\langle m \rangle)}{\Phi^{2}_{h}}, \ {\rm for} \ 1/2\leq \Phi_{h}<1,
    \label{qhalf}
\end{equation}
where $q_{1/2}$ stands for the cheating probability at which half of
the total money is owned by the cheaters, and half by the honest
players. In Fig.~\ref{Fig4}a we indeed see that for cheating probabilities beyond $q_{1/2}$ (blue dashed line) we find $\langle M_{c} \rangle>\langle M_{h} \rangle$. This finding has significant implications, as a small group of
cheaters can accumulate the majority of wealth by cheating with a
probability $q_{c}>q_{1/2}$, despite the fact that cheaters are also deceiving
each other. Remarkably, a single cheater can accumulate more wealth
than all honest players combined by cheating with a probability
$q_{c}>q_{1/2}$, as illustrated in Fig.~\ref{Fig4}a.

The threshold
value \eqref{qhalf} reaches a maximum in the low density limit of cheaters where $\Phi_{h}\rightarrow1^{-}$, which allows us to
establish a more general bound $q_{c}>(1/2)(\sqrt{4+\langle m \rangle^{2}}-\langle m \rangle)$.
This bound does \emph{not} require any prior
knowledge about the fraction of honest players, and therefore can be
utilized by a single cheater without any knowledge about the rest of the
players. In Fig.~\ref{Fig4}c this bound is shown with the blue dashed line. 

Finally, note that as $\langle m \rangle$
increases, the threshold value decreases, following an
asymptotic scaling 
\begin{equation} 
q_{1/2}=\frac{\Phi_{h}-1/2}{2\langle m
  \rangle}+\mathcal{O}\left(\frac{1}{\langle m \rangle^{2}}\right).
\end{equation}
This scaling corroborates the findings presented in
Sec.~\ref{SecIII.E}. Therefore, when $\langle m \rangle$ is large, a
single cheater only needs to cheat very little before
accumulating more than half of the total wealth. 
\subsection{When is cheating beneficial for a small pool of cheaters?}\label{SecIV.B}
The inset in Fig.~\ref{Fig4}a illustrates that there is a certain
lower range of cheating probabilities where $\langle M_{c} \rangle/M$
remains almost constant with respect to $q_{c}$. To explore this phenomenon
further, let us consider a small density of cheaters. We
can write $\Phi_{c}=\varepsilon$ with $0<\epsilon\ll1$ and consider an expansion of
$\pi^{h}(0)$ in $\varepsilon$, which is explicitly shown in Eq.~\eqref{seriessol}. Inserting this result for $\pi^{h}(0)$ into
Eq.~\eqref{mc}, we compute the total average wealth owned by the small
pool of cheaters 
\begin{equation}
    \Phi_{c}\langle m_{c} \rangle = \varepsilon \langle m_{c} \rangle = 
    \begin{dcases*}
            \varepsilon\frac{1-q_{\rm crit}^{2}}{2(q_{\rm crit}-q_{c})}+\mathcal{O}(\varepsilon^{2}), & $q_c<q_{\rm crit}$,\\ 
            \varepsilon^{1/2}\frac{(1-q^4_c)^{1/2}}{2q_c}-\varepsilon \frac{q_c(1+3q^{2}_{c})}{4(1+q^{2}_{c})}+\mathcal{O}(\varepsilon^{3/2}), & $q_c=q_{\rm crit}$,\\ 
            \langle m \rangle - \frac{1-q_{c}^{2}}{2q_{c}}+\varepsilon \frac{(\langle m \rangle+q_c)(1-q^{2}_c)}{q^{2}_{c}+2\langle m \rangle q_{c}-1}+\mathcal{O}(\varepsilon^{2}), & $q_{c}> q_{\rm crit}$, 
    \end{dcases*}
    \label{gain}
\end{equation}
where we identify a critical cheating probability
$q_{\rm crit}$ given by
\begin{equation}
    q_{\rm crit}\equiv \sqrt{\langle m \rangle^{2}+1}-\langle m \rangle,
    \label{qcrit}
\end{equation}
which is shown in Fig.~\ref{Fig4}c with the black dashed line. Note that $q_{\rm crit}$ coincides with Eq.~\eqref{sol2}, i.e., the probability for an honest player to have zero money in the game without cheaters. The reason for this will be explained in Sec.~\ref{SecIV.C}. From Eq.~\eqref{gain} we see that for $q_c< q_{\rm crit}$ we find that the cheaters
do \emph{not} substantially gain from cheating, as $\Phi_c \langle m_c
\rangle$ is proportional to the density of cheaters, $\varepsilon$. Similarly, at $q_c=q_{\rm
  crit}$ the wealth is proportional to $\varepsilon^{1/2}$. On the
contrary, for $q_c>q_{\rm crit}$ the first term in Eq.~\eqref{gain} is
independent of $\varepsilon$, leading to substantial gains. Taking the
limit $\varepsilon\rightarrow 0^{+}$ (corresponding to an arbitrarily
small density of cheaters), Eq.~\eqref{gain} reduces to 
\begin{equation}
    \lim_{\varepsilon\rightarrow0^{+}} \varepsilon \langle m_{c} \rangle = 
    \begin{dcases*}
            0, & $q_c\leq q_{\rm crit}$,\\ 
            \langle m \rangle - \frac{1-q_{c}^{2}}{2q_{c}}, & $q_{c}> q_{\rm crit}$, 
    \end{dcases*}
    \label{gain2}
\end{equation}
which is shown in Fig.~\ref{Fig4}b with the black solid line.  Therefore, when the pool of
cheaters vanishes with respect to the pool of honest players (i.e., when
$\Phi_c\rightarrow0^{+}$), cheating is only beneficial when
$q_{c}>q_{\rm crit}$. In particular, this includes the regime with a
finite number of cheaters amidst an infinite number of honest
players. Notably, at $q_c=q_{\rm crit}$ the total wealth of cheaters
undergoes a second-order discontinuity (i.e.\ a second-order phase transition), as the derivative of $\langle
m_{c} \rangle$ w.r.t.~$q_c$ is discontinuous at $q_c=q_{\rm crit}$ 
\begin{equation}
    \lim_{\varepsilon\rightarrow0^{+}} \varepsilon \frac{{\rm d}\langle m_{c} \rangle}{{\rm d}q_{c}} = 
    \begin{dcases*}
            0, & $q_c\leq q_{\rm crit}$,\\ 
            \frac{1}{2}\left(1+\frac{1}{q^{2}_{c}}\right), & $q_{c}> q_{\rm crit}$.
    \end{dcases*}
\end{equation} 
\subsection{Interpretation of the critical cheating probability}\label{SecIV.C}
The intuition behind the second-order discontinuity in Eq.~\eqref{gain2} can be understood from the
transition probability for a single cheater to gain and lose money in an infinite
pool of honest players. If we assume that the pool of honest players is
equilibrated in the absence of the cheater, the probability for an honest
player to have zero money is given by Eq.~\eqref{sol2}, which is identical to Eq.~\eqref{qcrit}, i.e., $\pi^h(0)=q_{\rm crit}$. We now introduce a single cheater with nonzero
wealth, and compute the transition probability $p^{c}(m+1|m)$ of gaining money upon encountering
a random honest player
\begin{equation} 
    p^{c}(m+1|m)=(1-\pi^{h}(0))[(1-q_{c})/2+q_c]=(1-\pi^{h}(0))(1+q_c)/2, \ m>0.
    \label{gainc}
\end{equation}
Similarly, the transition probability $p^c(m-1|m)$ for the single cheater to lose money to the honest player reads
\begin{equation}
    p^{c}(m-1|m)=\pi^{h}(0)(1-q_{c})+(1-\pi^{h}(0))(1-q_{c})/2=(1+\pi^{h}(0))(1-q_c)/2, \ m>0.
    \label{lossc}
\end{equation}
The reason why $\pi^{h}(0)$ plays such a key role in the transition probabilities is because a cheater cannot gain from a honest player with no money. Comparing Eqs.~\eqref{gainc} and \eqref{lossc} we see that the
probability of losing and gaining money are equal at $q_{c}=\pi^{h}(0)$. For $q_{c}<\pi^{h}(0)$ the cheater is more likely to lose
money than gain (i.e., $p^{c}(m-1|m)>p^{c}(m+1|m)$), and for $q_{c}>\pi^{h}(0)$ it is more likely for
the cheater to gain money (i.e., $p^{c}(m+1|m)>p^{c}(m-1|m)$). This effectively explains why $q_{\rm crit}=\pi^{h}(0)$ sets a
sharp transition between a flat and increasing trend for $\langle M_{c}\rangle$ as seen in Fig.~\ref{Fig4}b.
\subsection{Speed limits for losing and gaining money}\label{SecV.D}
In our previous analysis, we focused on the steady-state distribution
of wealth among cheaters and honest players. In this section, we shift
our focus to the dynamics of how the system relaxes toward this steady
state. By recognizing that the combined wealth of both cheaters and
honest players in our all-to-all game can be modeled as a discrete-time one-dimensional random
walk, we derive in Appendix \ref{AppendixD} an expression for the average wealth owned by the cheaters after $n$ encounters, which reads
\begin{equation}
    \langle M_{c}(n) \rangle = \langle M_{c}(0) \rangle+2 \Phi_{c}(1-\Phi_c)\sum_{l=0}^{n-1}(P^{c}(0;l)[1-q_{c}]-P^{h}(0;l)+q_{c}), \label{msol1}
\end{equation}
where $\langle M_{c}(0) \rangle$ is the initial wealth owned by the
cheaters. Furthermore, $P^{c}(0;n)$ and $P^{h}(0;n)$ are the probabilities after $n$
encounters for a cheater and honest player to have zero money,
respectively. Note that the average wealth owned by the honest players
follows from $\langle M_{h}(n) \rangle = M-\langle M_{c}(n)
\rangle$. 
\begin{figure}
    \centering
    \includegraphics[width=0.9\textwidth]{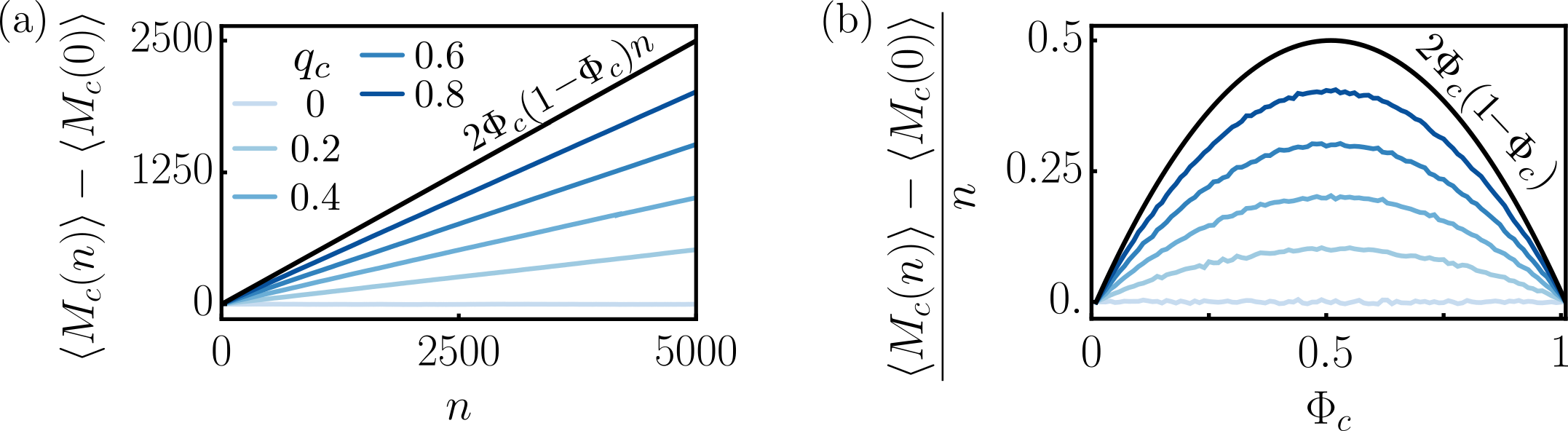}
    \caption{(a) Temporal evolution of the total wealth $\langle M_c
      (n) \rangle$ owned by the cheaters as a function of the number
      of encounters $n$ between two players. Colored lines are
      obtained by Monte-Carlo simulations with $N_h=50$ honest players,
      $N_{c}=50$ cheaters, and $M=10^{4}$ money units. The black solid
      line corresponds to the upper bound in Eq.~\eqref{mbound1}. (b)
      Rate of gaining money as a function of the fraction of
      cheaters; for each data point we evaluated the total wealth of
      the cheaters after $n=500$ steps. Colored lines are obtained by
      Monte-Carlo simulations with varying number of cheaters in the
      range $N_{c}\in\{0,...,100\}$, while keeping the total number of
      players fixed at $N=100$.  
    }
    \label{Fig5}
\end{figure}
Since $0\leq P^{c}(0;l) \leq 1$ and $0\leq P^{h}(0;l) \leq
1$, we can use the maximum and minimum values of both probabilities to
derive the following bounds for losing and gaining money after $n$
encounters 
\begin{equation}
    -2(1-q_{c}) \Phi_{c}(1-\Phi_{c})n \leq\langle M_{c}(n) \rangle-\langle M_{c}(0) \rangle \leq 2\Phi_{c}(1-\Phi_{c})n. \label{mbound1} 
\end{equation}
The upper bound is shown in Fig.~\ref{Fig5}a with the black line, and
saturates when $q_{c}\rightarrow1$. Notably, when $q_{c}=1$, the lower
bound in Eq.~\eqref{mbound1} equals $0$, since a cheater who always
cheats cannot incur losses. Most interestingly, at $\Phi_{c}=1/2$ the
lower and upper bound attain their largest value (in magnitude), which is also shown in
Fig.~\ref{Fig5}b. This shows that the fastest profit for a group of
cheaters occurs when half of the players are cheaters and the other
half are honest players. As the number of cheaters increases (i.e.,
$\Phi_c>1/2$), the likelihood of encountering an honest player
diminishes, which in turn reduces the rate at which cheaters can
profit. Conversely, a smaller number of cheaters (i.e., $\Phi_c<1/2$)
leads to diminished gains, as there are fewer cheaters that contribute
money to the cheaters pool.  
 
Since $\langle M_{h}(n)\rangle-\langle M_{h}(0)\rangle=\langle
M_{c}(0)\rangle-\langle M_{c}(n)\rangle$, the upper and lower bounds
in Eq.~\eqref{mbound1} are interchanged and multiplied by a negative
sign for the honest players. Consequently, it follows that when $q_{c} =
1$, the group of honest players cannot gain any money, which aligns with
our expectations. 
\section{Discussion \& Conclusion}\label{SecV}
We investigated the transient and steady-state dynamics
of the BDY game in the presence of probabilistic cheaters. The addition of probabilistic cheaters results in two groups of players each having an independent exchange strategy, resulting in different steady-state wealth distributions for each group. We derived analytical expressions for the steady-state
distribution of wealth for a single cheater and an honest player in
Sec.~\ref{SecIII}, and analyzed how those distributions and their moments depend on the model parameters such as the fractions of cheaters and honest players, the cheating probability, and the total wealth. In Sec.~\ref{SecIII.E} we found a critical scaling of the
cheating probability, $q_{c}\propto 1/\langle m \rangle$ for $\langle
m \rangle \gg 1$, which induces significant changes in the steady-state
distribution for the cheaters. This implies that already a very low
probability of cheating can have drastic consequences for the
honest players. Furthermore, in Sec.~\ref{SecIII.F} we derived a lower bound for the relative
variance of the combined wealth distribution, which can be used by an
external observer to detect the presence of hidden cheaters. A violation of this lower bound is a direct consequence of the addition of two different wealth distributions, namely that of the honest players and cheaters, which results in a bi-exponential distribution. Such a bi-exponential distribution is lacking in the original BDY game, where only honest players are present. 

In Sec.~\ref{SecIV} we determined the total average wealth held by
the cheaters and the honest players. When the fraction of cheaters is
lower than that of the honest players, i.e., $\Phi_c<\Phi_h$, we
identified a threshold cheating probability $q_{1/2}$ given by Eq.~\eqref{qhalf} above which the
smaller pool of cheaters will own the majority of money. Taking an
arbitrarily small density of cheaters, we also identified a critical
cheating probability $q_{\rm crit}$ given by Eq.~\eqref{qcrit} at which the average wealth of the
cheaters undergoes a second order discontinuity. From the perspective
of the cheaters it is beneficial to attain a cheating probability
above $q_{\rm crit}$ in order to gain substantially from the
cheating. Finally, we also determined the relaxation dynamics of the
average wealth in Sec.~\ref{SecV.D}, and found limits on how fast a group of cheaters and
honest players can gain or lose money. The fastest gain for the
cheaters is set when the density of cheaters is kept at $\Phi_c=1/2$,
a result confirmed by Monte-Carlo simulations. 

The wealth distributions predicted by our model are either exponential or bi-exponential, thus having exponential tails. While the exponential form often accurately describes the bulk of real-world wealth distributions \cite{RevModPhys.81.1703}, these also often display power-law tails. Within the BDY-type models, one possible explanation of the power-law statistics is that the exchange rules are not fair. Indeed, a power-law distribution has been derived for such an extended model with biased exchanges \cite{10.1088/1751-8121/ad369b}. How cheaters change the wealth distributions for the models with unfair exchange rules is an open question. 

Finally, we note that, while cheating
goes unpunished in our model, it would be interesting to explore the introduction of penalties for cheating, such as financial
repercussions upon detection. This approach would allow for a deeper
investigation of the optimal probabilities of detection and
cheating. Additionally, it would be valuable to categorize different
groups of cheaters, each with unique cheating probabilities, to
understand the dynamics of cheating behavior more
comprehensively. Such advancements could enhance our understanding of
the interplay between detection mechanisms and deception strategies.\vspace{0.5cm}\\ 
\\
\noindent \textbf{Acknowledgments.} Financial support from the German Research Foundation
(DFG) through the Heisenberg Program Grants GO 2762/4-1 and GO
2762/5-1 (to AG) is gratefully acknowledged. DEM is grateful to the support from the Robert A. Welch Foundation (Grant F-1514), US National Science Foundation (grant CHE-2400424), and the Alexander von Humboldt Foundation. 

\section*{Appendix}
\appendix
\renewcommand{\theequation}{A\arabic{equation}}
\renewcommand{\thefigure}{A\arabic{figure}}
\setcounter{equation}{0}
\setcounter{figure}{0}
\renewcommand{\theequation}{A\arabic{equation}}
\setcounter{equation}{0}
\section{Detailed balance in the money game}\label{AppendixA}
\begin{figure}
    \centering
    \includegraphics[width=0.95\textwidth]{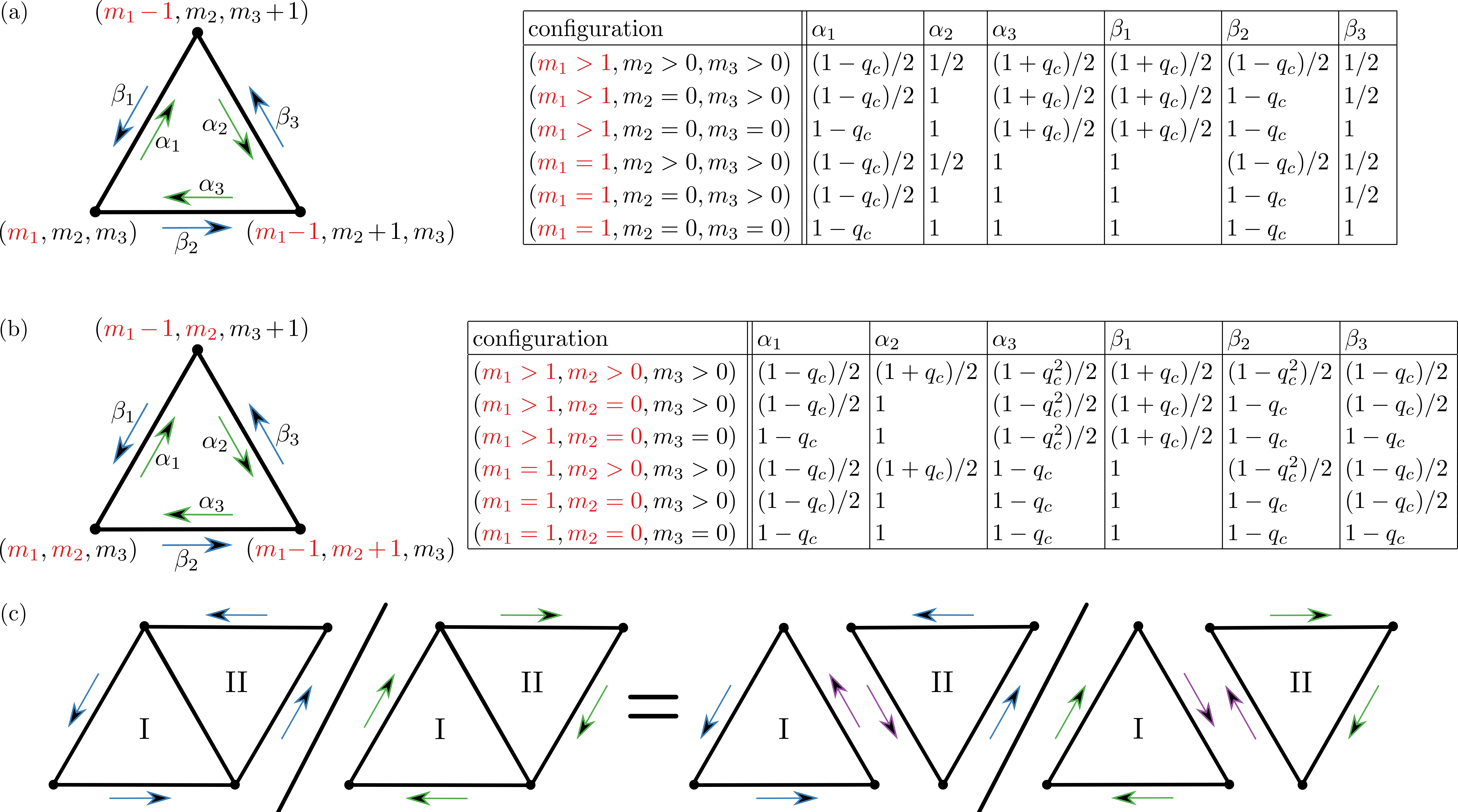}
    \caption{(a,b) Transition probabilities (up to a multiplicative
      constant factor) for the exchange of money between (a) one
      cheater and two honest players, and (b) two cheaters and one honest
      player. The amount of money owned by the cheaters is indicated
      in red. For each possible wealth scenario between the players,
      the table on the right shows that the product of clockwise to
      counterclockwise transition probabilities is equal. This
      confirms that the money game with cheaters obeys detailed
      balance. (c) Utilizing the exchange triangles, we can derive the
      ratio of clockwise to counterclockwise transition probabilities
      for a diamond graph (and any larger graph). It is important to
      note that the transition probabilities highlighted in purple
      cancel out in the ratio, ensuring that they are not overcounted.} 
    \label{Fig6}
\end{figure}
To demonstrate that the money game involving cheaters satisfies
detailed balance (see Eq.~\eqref{DB}), we utilize Kolmogorov’s cycle
criterion \cite{kolmogoroff1936theorie,kelly2011reversibility}. This
criterion establishes that detailed balance is satisfied if and only
if all cyclic sequences of states  
have an equal product of transition probabilities in the clockwise and
counterclockwise direction, respectively.   

In the context of the money game, we leverage the fact that the
connectivity graph between all wealth configurations can be
constructed from exchange triangles involving three players, as
illustrated in Fig.~\ref{Fig6} and Fig.~\ref{Fig2}d-f. Specifically,
Fig.~\ref{Fig6}a depicts an exchange triangle that showcases the
potential exchanges between a cheater and two honest players. The
accompanying Table in Fig.~\ref{Fig6}a demonstrates that for all
possible wealth distributions among the players, the relationship
$\alpha_1\alpha_2\alpha_3 = \beta_1\beta_2\beta_3$ holds, where
$\alpha_i$ and $\beta_i$ denote the transition probabilities in the
counterclockwise and clockwise directions, respectively. This finding
also extends to exchanges involving two cheaters and one honest player,
as shown in Fig.~\ref{Fig6}b. For exchanges involving three cheaters or
three honest players, the symmetry of the exchange triangle inherently
ensures equality between a cycle of clockwise and counterclockwise transitions.  

From these exchange triangles, we can derive the ratio of clockwise to
counterclockwise transition probabilities for any larger cycle. For
example, in Fig.~\ref{Fig6}c, we demonstrate how the transition
probability ratio in a diamond graph can be decomposed into the
transition probability ratios of the exchange triangles forming that
diamond structure. Together with the results presented in the Tables
of Fig.~\ref{Fig6}a,b, this analysis indicates that the ratio of
clockwise to counterclockwise transition probabilities is equal to one
for any cycle. Thus, we conclude that the money game with cheaters
obeys detailed balance. 
\renewcommand{\theequation}{B\arabic{equation}}
\setcounter{equation}{0}
\section{Derivation of the steady-state wealth distribution per player}\label{AppendixB}
\subsection{Steady-state probability distribution of wealth for a cheater}
We concentrate on a single cheater and examine the evolution of the
wealth $m$ owned by this individual. We employ the
local-equilibrium approximation \cite{PhysRevX.11.031067,
  PhysRev.148.375}, assuming that the dynamics of an individual player
can be effectively modeled by a one-dimensional Markovian random walk with effective transition probabilities of stepping right and left estimated as averages over the rest of the players.
Our objective is to evaluate, $p^{c}(m+1|m)$ and $p^{c}(m-1|m)$,  the conditional
probabilities that a cheater with an amount $m$ of money will gain or lose money
upon interacting with another player. Focusing on $p^{c}(m+1|m)$, we note that the cheater
can only gain when the other player has nonzero wealth. This event can only result from the following scenarios: 
 \begin{itemize}
     \item the other player is an honest player and has nonzero money: the probability of this is $\Phi_{h}(1-\pi^{h}(0))$
     \item the other player is a cheater, does not cheat, and has nonzero money: the probability is $\Phi_{c}(1-q_{c})(1-\pi^{c}(0))$
 \end{itemize}
Now we also need to consider whether the cheater is actually cheating in a respective move or not. 
Taking this into account,  the conditional jump probability for a cheater to gain reads
 \begin{align}
     p^{c}_{+}\equiv p^{c}(m+1|m)&=(2/N)[\Phi_{h}(1-\pi^{h}(0))+\Phi_{c}(1-q_{c})(1-\pi^{c}(0))][q_{c}+(1-q_{c})(1/2)], \ m\geq 1.
     \label{pc1}
 \end{align} 
In the above equation, the factor $2/N$ accounts for the probability of selecting a
specific player, which is $1/N$ times $2$ since the encounter between
two players can be initiated from both sides. Note that this factor
will appear in all conditional jump probabilities.  It therefore
cancels whenever a  ratio of probabilities is taken. Next, we also need to consider
separately the case when the cheater has zero money, for which a
cheating or fair move are equivalent, resulting in 
\begin{equation}
     \hat{p}^{c}_{+}\equiv p^{c}(1|0)=(2/N)[\Phi_{h}(1-\pi^{h}(0))+\Phi_{c}(1-q_{c})(1-\pi^{c}(0))].
     \label{pc2}
 \end{equation}
Finally, we consider the conditional probability to lose money, i.e., $p^{c}(m-1|m)$. Note that the cheater can only lose money upon not cheating. Taking into account that the other player can be a cheater or an honest player, this leads to
\begin{align}
      p^{c}_{-}\equiv p^{c}(m-1|m)&=(2/N)(1{-}q_{c})\!\left(\Phi_{h}[\pi^{h}(0){+}(1{-}\pi^{h}(0))/2]{+}\Phi_{c}[\pi^{c}(0){+}(1{-}\pi^{c}(0))\left(q_c{+}(1{-}q_c)/2\right)]\right), \ m\geq 1.
      \label{pc3}
\end{align}
Since the dynamics of a single cheater in the money game takes place
on a one-dimensional graph, the steady-state probabilities
$\pi^{c}(m)$ automatically must satisfy the detailed balance condition
\cite{norris1998markov} (this is true regardless of whether the full
underlying Markov chain obeys detailed balance or not), i.e., 
\begin{equation}
\pi^{c}(m) p^{c}(m\pm 1|m)=\pi^{c}(m\pm 1)p^{c}(m|m\pm 1).  
 \label{detailed balance}
\end{equation}
Equation~\eqref{detailed balance} can be used as an iterative map to obtain $\pi^{c}(m)$, starting from $\pi^{c}(0)$
\begin{equation}
    \pi^{c}(1)=\pi^{c}(0)\frac{\hat{p}^{c}_{+}}{p^{c}_{-}}, \ \pi^{c}(2)=\pi^{c}(0)\frac{p^{c}_{+}\hat{p}^{c}_{+}}{(p^{c}_{-})^{2}}, \ {\rm etc.},
\end{equation}
and since the conditional transition probabilities do not depend explicitly on $m$, we obtain the general result
\begin{equation} 
\pi^{c}(m) = 2\pi^{c}(0)\frac{(1+q_{c})^{m-1}[\Phi_{h}(1-\pi^{h}(0))+\Phi_{c}(1-q_{c})(1-\pi^{c}(0))]^{m}}{(1-q_{c})^{m}[\Phi_{h}(1+\pi^{h}(0))+\Phi_{c}(2-(1-q_{c})(1-\pi^{c}(0)))]^{m}}, \ m\geq1.
\label{pic1}
\end{equation}
Note that both, $\pi^{c}(0)$ and $\pi^{h}(0)$ still need to be determined. To determine $\pi^{h}(0)$ we use the normalization condition
\begin{equation}
\sum_{m=0}^{\infty} \pi^{c}(m)  = 1,
\end{equation}
and obtain the remarkably simple result given by Eq.~\eqref{solc}. Inserting this expression for $\pi^{h}(0)$ into Eq.~\eqref{pic1} finally results in Eq.~\eqref{pic2}. 
To determine the expression for $\pi^{h}(0)$, we need to evaluate the
steady-state of the honest players, which we do in the next section. 
\subsection{Steady-state probability distribution of wealth for an honest player}
We now follow the same procedure as described for the cheaters but
instead focus on the random walk of a single honest player in the
game. Let $p^{h}(m+1|m)$ and $p^{h}(m-1|m)$ be the conditional
probability that an honest player with $m$ units of money will increase/decrease
said amount by $1$. Using the same reasoning as we did for the
conditional probabilities of the cheater, we obtain
 \begin{alignat}{2}
     \!p^{h}_{+}&\equiv p^{h}(m+1|m)=(2/N)[\Phi_{h}(1-\pi^{h}(0))+\Phi_{c}(1-q_{c})(1-\pi^{c}(0))]/2 &&=(2/N)(1-\pi^{h}(0))/2, \ m\geq 1, \label{pf1}  \\
     \!\hat{p}^{h}_{+}&\equiv p^{h}(1|0)=(2/N)[\Phi_{h}(1-\pi^{h}(0))+\Phi_{c}(1-q_{c})(1-\pi^{c}(0))] &&=(2/N)(1-\pi^{h}(0)), \label{pf2} \\
     \!p^{h}_{-}&\equiv p^{h}(m-1|m)=(2/N)[\Phi_{h}(1+\pi^{h}(0))
     +\Phi_{c}(2-(1-q_{c})(1-\pi^{c}(0)))]/2 
     &&=(2/N)(1+\pi^{h}(0))/2, \ m\geq 1, \label{pf3}
 \end{alignat}
 where for the last equalities we have inserted Eq.~\eqref{solc} for $\pi^{c}(0)$. Remarkably, we obtain the same expressions for the conditional probabilities as for the game with no cheaters (see Eqs.~(4) and (5) in \cite{10.1088/1751-8121/ad369b}). However, note that $\pi^{h}(0)$ will depend on the presence of cheaters. To see this, let us first construct the steady-state probability for the honest players, which can be obtained from the detailed balance relation
\begin{equation}
     \pi^{h}(m)p^{h}(m\pm1|m)=\pi^{h}(m)p^{h}(m|m\pm1).
     \label{DBf}
\end{equation}
Upon iterating Eq.~\eqref{DBf} for $\pi^{h}(m)$ we obtain Eq.~\eqref{pifm}. Notably, Eq.~\eqref{pifm} is already normalized, i.e., $\sum_{m=0}^{\infty}\pi^{h}(m)=1$, regardless of the value of $\pi^{h}(0)$. So, to determine $\pi^{h}(0)$, we need to use the conservation of money $\langle m \rangle$, which reads
 \begin{equation}
     \Phi_{h}\langle m_{h} \rangle +\Phi_{c} \langle m_{c} \rangle =\langle m \rangle,
     \label{constr}
 \end{equation}
 where $\langle m_{c} \rangle$ and $\langle m_{h} \rangle$ are given by Eqs.~\eqref{mc}-\eqref{mf}.
 Inserting the expressions for $\langle m_{c} \rangle$ and $\langle m_{h} \rangle$ into Eq.~\eqref{constr}, and using that $\Phi_{c}=1-\Phi_{h}$, we obtain after some algebraic manipulations
 \begin{equation}
       [\pi^{h}(0)]^{3}+[\pi^{h}(0)]^{2}(2\langle m \rangle -\Phi_{h}q_{c})-\pi^{h}(0)(1+2\langle m \rangle q_{c})+\Phi_{h}q_{c}=0.
       \label{cond2}
\end{equation} 
By writing 
\begin{equation}
    \pi^{h}(0)=t+(1/3)(\Phi_{h}q_{c}-2\langle m \rangle),
\end{equation}
we can transform Eq.~\eqref{cond2} into a suppressed cubic for $t$
\begin{equation}
    t^{3}-\Delta_0 t+\Delta_1=0,
\end{equation}
where $\Delta_0$ and $\Delta_1$ are given by Eq.~\eqref{del1}, and whose solution is given by Eq.~\eqref{solf}. This concludes the derivation of the steady-state wealth distribution for a cheater and honest player.  
\subsubsection{Perturbation expansion for $\Phi_c=\varepsilon$}
Here we consider $\Phi_{c}=\varepsilon \ll 1$ (such that
$\Phi_h=1-\varepsilon$ is close to $1$), allowing us to use a perturbation expansion of Eq.~\eqref{cond2}. The leading
order solutions of Eq.~\eqref{cond2} for $\Phi_h=1$ in the interval
$[0,1]$ are given by $\pi^h(0)=q_{\rm crit}$ (see Eq.~\eqref{sol2})
and $\pi^h(0)=q_c$. Since ${\rm max}(q_{c},q_{\rm
  crit})<\pi^{h}(0)\leq1$ (see proof in Appendix \ref{AppendixC}),
we choose the leading order term $\pi^h(0)=q_{\rm crit}$ for  $q_c
\leq q_{\rm crit}$, and $\pi^h(0)=q_c$ for $q_c>q_{\rm crit}$. Next,
we compute the first order perturbation. Here, special care has to be
taken when $q_c=q_{\rm crit}$, as the first order perturbation here scales
with $\varepsilon^{1/2}$ instead of $\varepsilon$. The
final result for $\pi^{h}(0)$ then reads 
\begin{equation}
    \pi^{h}(0)\simeq
    \begin{dcases*}
       q_{\rm crit}+\varepsilon\frac{\langle m \rangle q_c}{\sqrt{\langle m \rangle ^{2}+1}(1-\langle m \rangle q_c)-q_{c}(1+\langle m \rangle^2)}+\mathcal{O}(\varepsilon^{2}), & $q_c<q_{\rm crit}$,\\ 
        q_{\rm crit}+\varepsilon^{1/2}\left(\frac{1-q^2_c}{1+q^2_c}\right)^{1/2}-\varepsilon\frac{q^{3}_{c}(3+q^{2}_{c})}{2(1+q^{2}_{c})^{2}}+\mathcal{O}(\varepsilon^{3/2}), & $q_c=q_{\rm crit}$,\\
        q_{c}+\varepsilon\frac{ q_{c}(1-q^2_c)}{q^2_c+2\langle m \rangle q_c-1} -\varepsilon^{2}\frac{2\langle m \rangle q^{2}_{c}(1-q^{4}_{c})}{(q^2_c+2\langle m \rangle q_c-1)^{3}}+\mathcal{O}(\varepsilon^{3}),& $q_c>q_{\rm crit}$.\
    \end{dcases*}
\label{seriessol}
\end{equation}
The first order terms in Eq.~\eqref{seriessol} are positive, and therefore we find ${\rm max}(q_{c},q_{\rm crit})<\pi^{h}(0)\leq1$ as expected. Inserting Eq.~\eqref{seriessol} into Eq~\eqref{mc}
results in the expansion shown in Eq.~\eqref{gain}. 
\renewcommand{\theequation}{C\arabic{equation}}
\setcounter{equation}{0}
\section{Proof of Eq.~\eqref{varineq} in the presence of cheaters}\label{AppendixC}
To prove the inequality given by Eq.~\eqref{varineq} we employ the following steps: 
\begin{enumerate}
    \item First, note that $\pi^{h}(0)$ obeys the the following bound for $0<\Phi_c\leq1$, $0<q_c\leq1$, and $\langle m \rangle>0$ 
    \begin{equation}
       q_{\rm crit} < \pi^{h}(0)\leq 1,
        \label{boundie}
    \end{equation}  
    where $q_{\rm crit}$ is given by Eq.~\eqref{sol2}. 
    To prove this, we substitute the value $\pi^{h}(0)\rightarrow
    q_{\rm crit}$ into
    Eq.~\eqref{cond2}, which leaves us with a negative remainder term
    of $-2\langle m \rangle q_{\rm crit}\Phi_{c}q_{c}< 0$. Similarly, if we substitute the value
    $\pi^{h}(0)\rightarrow 1$ into Eq.~\eqref{cond2}, we obtain a
    nonnegative remainder term of $2\langle m \rangle (1-q_c)$.  This implies that somewhere in the interval $[q_{\rm crit},1]$ the cubic equation \eqref{cond2} crosses the value $0$, from which follows Eq.~\eqref{boundie}. Furthermore, together with the results from Sec.~\ref{SecIII}, this implies that ${\rm max}(q_{c},q_{\rm crit})<\pi^{h}(0)\leq1$. 
    \item Second, starting from Eq.~\eqref{constr}, we can rewrite Eq.~\eqref{cond2} into the following form
    \begin{equation}
        \frac{1}{\pi^{h}(0)}+\frac{(1-[\pi^{h}(0)]^2)\Phi_c q_{c}}{(\pi^{h}(0)-q_{c})\pi^{h}(0)}=2\langle m \rangle+\pi^{h}(0).
        \label{rewrite}
    \end{equation}
    The left hand side of  Eq.~\eqref{rewrite} is similar to
    the expression inside the brackets of Eq.~\eqref{varm}. The right
    hand side of Eq.~\eqref{rewrite} can be bounded from below using Eq.~\eqref{boundie}
    \begin{equation}
        2\langle m \rangle +\pi^{h}(0)\stackrel{\eqref{boundie}}{>} 2\langle m \rangle + q_{\rm crit}= \langle m \rangle+\sqrt{\langle m\rangle ^{2}+1}.
        \label{boundieresult}
    \end{equation}
    \item Finally, starting from Eq.~\eqref{varm}, we use the following chain of (in)equalities 
    \begin{align}
        \frac{{\rm var}(m)}{\langle m\rangle^{2}}&\!\!\stackrel{\eqref{varm}}{=}\frac{1}{\langle m \rangle}\left(\frac{1}{\pi^{h}(0)}+\frac{(1-[\pi^{h}(0)]^2)\Phi_c q_{c}}{(\pi^{h}(0)-q_{c})(\pi^{h}(0)-\Phi_hq_c)}\right)-1 \nonumber \\
        &\geq \frac{1}{\langle m \rangle}\left(\frac{1}{\pi^{h}(0)}+\frac{(1-[\pi^{h}(0)]^2)\Phi_c q_{c}}{(\pi^{h}(0)-q_{c})\pi^{h}(0)}\right)-1 \nonumber \\
        &\!\!\stackrel{\eqref{rewrite}}{=}\frac{1}{\langle m \rangle}\left(2\langle m \rangle +\pi^{h}(0)\right)-1 \nonumber \\
        &\!\!\stackrel{\eqref{boundieresult}}{>}\sqrt{1+1/\langle m \rangle^{2}},
    \end{align}
    where the first inequality follows from $1/(x-a)\geq 1/x$ for
    $x\geq a\geq0$. This concludes the proof of Eq.~\eqref{varineq}.   
    
\end{enumerate}
\renewcommand{\theequation}{D\arabic{equation}}
\setcounter{equation}{0}
\section{Relaxation dynamics}\label{AppendixD}
Here, we derive Eq.~\eqref{msol1} for the transient
dynamics of the money owned by the cheaters. Our starting point is a
master equation for the temporal evolution of the
money
distributions for a single cheater and honest player, denoted as
$P^{c}(m;n)$ and $P^{h}(m;n)$, respectively. Note that by definition,
$\lim_{n\rightarrow \infty}P^{c}(m;n)=\pi^{c}(m)$ and
$\lim_{n\rightarrow \infty}P^{h}(m;n)=\pi^{h}(m)$. The master equation
for the probability distribution of a single cheater reads 
\begin{align}
    P^{c}(0;n+1)&=p^{c}_{-}(n)P^{c}(1;n)-(\hat{p}^{c}_{+}(n)-1)P^{c}(0;n) \nonumber , \\ 
    P^{c}(1;n+1)&=p^{c}_{-}(n)P^{c}(2;n)+\hat{p}^{c}_{+}(n)P^{c}(0;n)-[p^{c}_{-}(n)+p^{c}_{+}(n)-1]P^{c}(1;n) \label{meq2}, \\ 
    P^{c}(m;n+1)&=\sum_{k=\pm}[p^{c}_{k}(n)P^{c}(m-k;n)-p^{c}_{k}(n)P^{c}(m;n)] +P^{c}(m;n), \ m\geq 2. \nonumber
\end{align}
For an honest player, the same results apply upon substituting
$c\rightarrow h$ in the above equations. Note that the transition
probabilities $p^{c}_{\pm}(n)$ and $\hat{p}^{c}_{+}(n)$ are
time-dependent, and are derived from Eqs.~\eqref{pc1}-\eqref{pc3} (and
Eqs.~\eqref{pf1}-\eqref{pf3} for the honest players) upon making the
substitution $\pi^{c}(0)\rightarrow P^{c}_{0}(n)$ and
$\pi^{h}(0)\rightarrow P^{h}_{0}(n)$,  i.e., the probability after $n$
steps for a cheater or honest player to have zero money, inside their respective
expressions. From this definition, it follows that
$\lim_{n\rightarrow\infty}p^{c}_{\pm}(n)=p^{c}_{\pm}$ and
$\lim_{n\rightarrow\infty}p^{h}_{\pm}(n)=p^{h}_{\pm}$. Note that the
master equations for $P^{h}(m;n)$ and $P^{c}(m;n)$ are coupled and
nonlinear in $n$ due to the nonlinearities in the transition probabilities $p^c_\pm(n)$ and $\hat{p}^c_\pm$(n). 
To solve Eq.~\eqref{meq2}, we introduce the generating
functions 
\begin{equation}
    G^{c}_{z}(n)\equiv \sum_{m=0}^{\infty}z^{m} P^{c}(m;n)
    \label{genF}
\end{equation}
Multiplying Eq.~\eqref{meq2} with $z^{m}$, and summing up all contributions, we obtain, after some algebraic manipulation,
\begin{equation}
    G^{c}_{z}(n+1)-G^{c}_{z}(n)=p^{c}_{-}(n)[G^{c}_{z}(n)-P^{c}(0;n)](z^{-1}-1){+}[p^{c}_{+}(n)(G^{c}_{z}(n)-P^{c}(0;n))+\hat{p}^{c}_{+}(n)P^{c}(0;n)](z-1) \label{Geq1}.
\end{equation}
Note that the right-hand side of Eqs.~\eqref{Geq1} is zero for $z=1$, which is due to the normalization condition $G^{c}_{1}(n)=\sum_{m=0}^{\infty}P^{c}(m;n)=1$. 
Expressing $\langle M_{c}(n) \rangle$ via the generating function. i.e.\ $\langle M_{c}(n) \rangle/N_{c}=dG^{c}_{z}(n)/dz|_{z=1}$, we obtain a difference equation for the average amount of money owned by the cheaters 
\begin{equation}
    \langle M_{c}(n+1)\rangle-\langle M_{c}(n)\rangle
    =2\Phi_{c}(1-\Phi_c)(P^{c}(0;n)[1-q_{c}]-P^{h}(0;n)+q_{c}) \label{dmct}.
\end{equation}
In general, one can obtain a difference equation for any arbitrary moment $\langle M_{c}^{k}(n) \rangle$ by differentiating the generating function $k$ times (see Eq.~\eqref{genF}).
The solution of Eq.~\eqref{dmct} is given by Eq.~\eqref{msol1}. Finally, note that upon inserting the steady-state relation
$\lim_{n\rightarrow\infty}P^{c}(0;n)=\lim_{n\rightarrow\infty}(P^{h}(0;n)-q_{c})/(1-q_{c})$
given by Eq.~\eqref{solc}, the right-hand side of Eq.~\eqref{dmct} becomes zero
as expected. 
%

\end{document}